\documentclass[reprint, superscriptaddress,nofootinbib]{revtex4-1}

\usepackage[breaklinks,colorlinks,
   urlcolor=blue,citecolor=blue,linkcolor=blue]{hyperref}
\usepackage[all]{hypcap}

\usepackage{amsfonts}
\usepackage{amssymb}
\usepackage{amsmath}

\usepackage{color}

\usepackage{graphicx}

\usepackage{lineno}

\newcommand{\pg}{\ensuremath{p\text{-}g}}

\newcommand{\Hnl}{\ensuremath{\mathcal{H}_{pg}}}
\newcommand{\Hgr}{\ensuremath{\mathcal{H}_{!pg}}}
\newcommand{\lnB}{\ensuremath{\ln B^{pg}_{!pg}}}
\newcommand{\logtenB}{\ensuremath{\log_{10} B^{pg}_{!pg}}}
\newcommand{\Bayes}{\ensuremath{B^{pg}_{!pg}}}

\newcommand{\Syr}{RB}
\newcommand{\LVC}{LVC}

\newcommand{\approxSyrSampSyrPrior}{\ensuremath{+0.7}}

\newcommand{\approxSyrBayes}{\ensuremath{+2.0}}

\begin{document}

\title{
A Comparison of $p$-$g$ Tidal Coupling Analyses
}

\author{Reed Essick}
\affiliation{Kavli Institute for Cosmological Physics, University of Chicago, \\Chicago, IL 60637, USA}

\author{Nevin N. Weinberg}
\affiliation{Department of Physics, and Kavli Institute for Astrophysics and Space Research, Massachusetts Institute of Technology, \\Cambridge, MA 02139, USA}

\begin{abstract}
Two recent studies have attempted to constrain the proposed \pg~tidal instability with gravitational-wave data from GW170817.
The studies use Bayesian methods to compare a model that includes \pg~tidal effects with one that does not.
Using the same data, they arrive at very different conclusions.
Reyes \& Brown find that the observations of GW170817 strongly disfavor the existence of $p$-$g$ mode coupling.
However, the LIGO and Virgo Collaborations find that neither model is strongly favored.
We investigate the origin of this discrepancy by analyzing Reyes \& Brown's publicly available posterior samples.  Contrary to their claims, we find that their samples do not disfavor \pg~mode coupling. 
\end{abstract}

\maketitle

\section{Introduction}
\label{section:introduction}

The gravitational wave (GW) observation of a coalescing binary neutron star (NS) system (GW170817 \cite{GW170817}) provides new insights into NS physics, including constraints on the high-density equation of state \cite{EOS, GW170817GRB} and tidal deformability \cite{De2018, GW170817SourceProperties, EOS}.
Recently, two papers have attempted to constrain the \pg~tidal instability with GW170817 (\cite{Reyes2018, LVCnltides}; hereafter \Syr~and \LVC, respectively).
The instability involves a non-resonant coupling of the linear tidal bulge to high-frequency, pressure-supported modes ($p$-modes) and low-frequency, gravity-supported modes ($g$-modes) within the NS \cite{Weinberg2013, Venumadhav2014, Weinberg2016, Zhou2017}.
Once unstable, the excited modes continuously drain energy from the orbit and accelerate the rate of GW-driven inspiral.
The precise impact on the phasing of the GW signal is, however, unknown due to theoretical uncertainties in how the instability grows and saturates, although studies suggest that its impact might be observable with the current LIGO \cite{LIGO} and Virgo \cite{Virgo} interferometers \cite{Weinberg2016, Essick2016}.

\Syr~and \LVC~attempt to constrain \pg~effects in GW170817 using the phenomenological model developed by \cite{Essick2016}.
Both studies employ a modification of the \texttt{TaylorF2} frequency-domain waveform (see, e.g., \cite{Buonanno2009}) that includes an additional phase correction induced by \pg~effects.
Using Bayesian methods, they compare models with \pg~effects (\Hnl) to models without \pg~effects (\Hgr) and compute Bayes Factors $\Bayes \equiv p(\mathrm{D}|\Hnl)/p(\mathrm{D}|\Hgr)$, where $D$ refers to the data from GW170817.

While \Syr~and \LVC~analyze the same data and use the same phenomenological \pg~waveform, there are differences in their models and priors, which we describe in Section~\ref{section:priors}.  Most notably,  \Syr~constructs an \Hnl~model that only includes ``detectable \pg~effects'' whereas \LVC~uses a wider \Hnl~model.  \Syr~finds that their models yield $\Bayes < 10^{-4}$ and \LVC~finds that their models yield $\Bayes \approx 1$.  Thus, \Syr~concludes that the observations strongly disfavor their \Hnl~model and \LVC~concludes that the observations do not favor either of their models.  

A priori, the disparate \Bayes~could be due to differences in the studies' models and priors.  However, we show in Section~\ref{section:bayes factors} that this cannot be the explanation.  We use the posterior samples from \Syr\footnote{\Syr~have kindly made their posterior samples available at \texttt{https://github.com/sugwg/gw170817-pg-modes}.} to compute $\Bayes$ using \LVC's method for calculating Bayes Factors \cite{LVCnltides}. We find that \LVC's method applied to \Syr's posterior samples, and thus their models and priors, yield $\Bayes \approx 1$ and not $\Bayes < 10^{-4}$.  This indicates that there is an error in how \Syr~calculates \Bayes.
Our estimate implies that their \Hnl~model is not disfavored by the data.

\section{Comparison of Models and Priors}
\label{section:priors}

The phenomenological model presented in \cite{Essick2016} introduces three  \pg~parameters per NS (indexed by $i\in\{1,2\}$): an overall amplitude ($A_i$) related to how many modes become unstable, how quickly they grow, and the energy at which they saturate; a turn-on/saturation frequency ($f_i$) that is related to when the modes first become unstable; and a spectral index ($n_i$) that describes how the rate of energy dissipation evolves with the orbital frequency (see  \cite{Essick2016}, \Syr, and \LVC).

The frequency-domain phase shift $\Delta \Psi(f)$ induced by \pg~effects is given by Equation (3) in \Syr~and Equation (1) in \LVC.
To account for a possible dependence on the component masses ($m_i$), \cite{Essick2016} introduces a Taylor expansion of the \pg~parameters around $m_i=1.4 M_\odot$.
\LVC~keeps the zeroth- and first-order coefficients of the expansion (their Equation (2)).
\Syr~keeps only the zeroth order coefficients.
However, this should not introduce large discrepancies since \LVC~and \cite{Essick2016} find that the first order terms are not measurable.

In their Equation (3), \Syr~neglects a dependence on the component masses that exists independent of the Taylor expansion.
Specifically, in the expression for $\Delta \Psi(f)$, there is a factor  $C_i=[2m_i/(m_1+m_2)]^{2/3}A_i$ and \Syr~assumes $C_1/A_1 = C_2/A_2$ even when $m_1 \neq m_2$.
However, since the difference is small for reasonable ranges of $m_i$, this should not introduce a large discrepancy.

\Syr's priors on the non-\pg~parameters are somewhat different from \LVC's.
\Syr~considers both a uniform and a Gaussian prior on $m_{i}$, whereas \LVC~considers only a uniform prior on $m_i$.
While \Syr's estimates of \Bayes~vary by as much as a factor of $10^3$ for different mass priors, they note that their posterior distributions are qualitatively very similar.
In addition, unlike \LVC, \Syr~assumes a fixed source location and distance based on the electromagnetic counterparts to GW170817 \cite{GW170817multimessenger,GW170817GRB}.
However, \cite{Essick2016} found that extrinsic parameters do not strongly impact the inference of \pg~effects.

\Syr~also assumes the NSs have equal radii and their priors on component spins differ from \LVC's.
However, we do not believe this could introduce large discrepancies.

The differences between \Syr's and \LVC's priors on the \pg~parameters are more substantial.
The most significant difference is that \LVC~assumes uniform priors on the zeroth order coefficients $\log_{10} A_0$, $f_0$, and $n_0$ (following \cite{Essick2016}), whereas \Syr~constrains the \pg~parameters to values that produce total time-domain phase shifts $\delta \phi \geq 0.1\, \mathrm{rad}$.
Thus, \Syr's priors on $\log_{10} A_0$, $f_0$, and $n_0$ are not uniform, but favor combinations that produce relatively large \pg~tidal effects.
\Syr~explains that for $m_1=m_2=1.4M_\odot$, there is a 99.98\% overlap between the waveforms from \Hnl~and \Hgr~for values of $(A_0,f_0, n_0)$ that yield $\delta \phi \approx 0.1\textrm{ rad}$.
As a result, like \LVC, their prior still allows certain limits of \Hnl~to reproduce \Hgr.

Finally, \LVC~focuses on a somewhat narrower bandwidth than \Syr~(minimum frequencies of 30 Hz vs. 20 Hz).
\LVC~does explore minimum frequencies above 30 Hz and find that $\Bayes$ only varies by factors of order unity (see their Figure 1).
Since the gain in signal-to-noise ratio from 30 Hz to 20 Hz is relatively modest ($\lesssim$ few percent), we do not believe this difference introduces large discrepancies.

\section{Comparison of results}
\label{section:bayes factors}

\begin{figure}
    \includegraphics[width=\columnwidth, clip=True, trim=1.0cm 0.0cm 1.5cm 7.7cm]{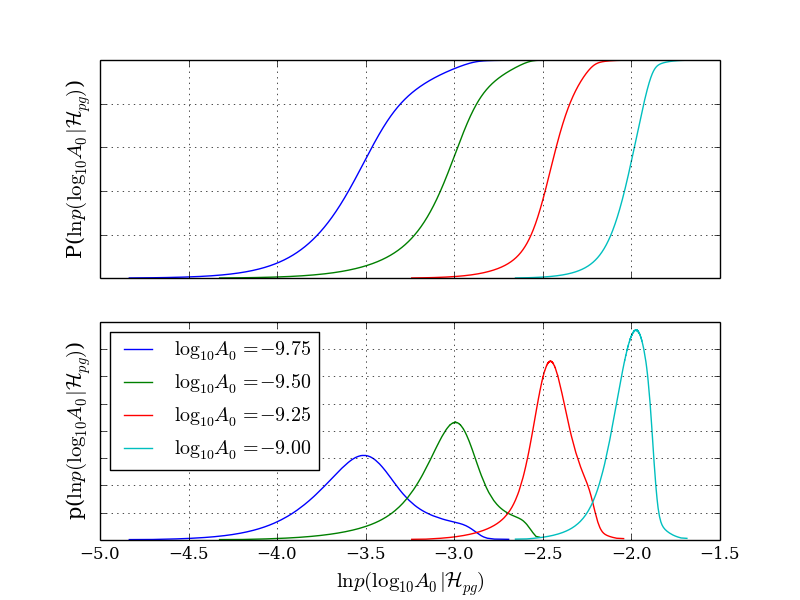}
    \includegraphics[width=\columnwidth, clip=True, trim=1.0cm 0.0cm 1.5cm 7.7cm]{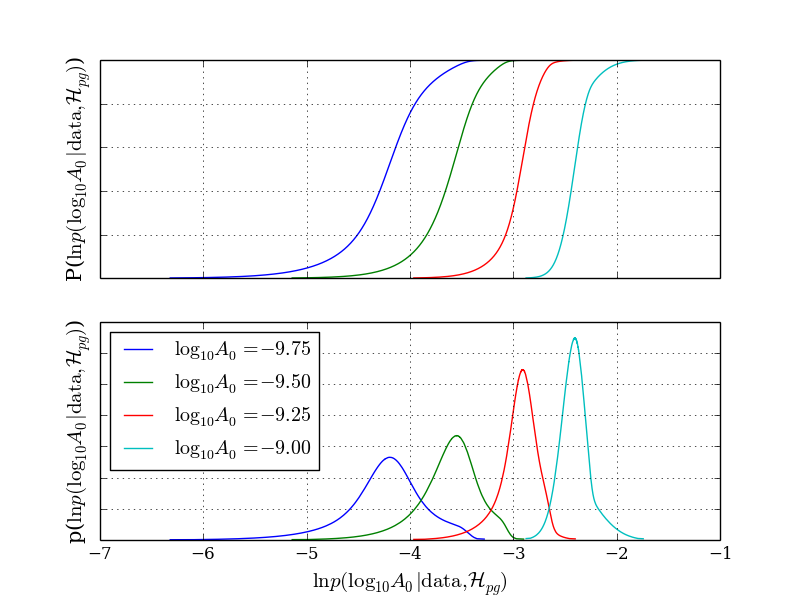}
    \includegraphics[width=\columnwidth, clip=True, trim=1.0cm 0.0cm 1.5cm 7.7cm]{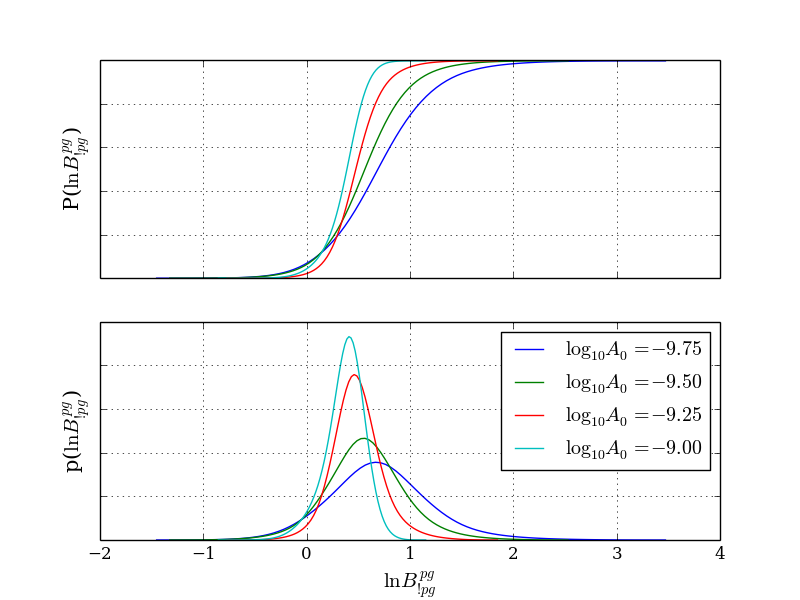}
    \caption{
        Posteriors, priors and \Bayes~computed using \Syr's posterior samples showing disagreement with \Syr's claim that $\Bayes < 10^{4}$.
        (\textit{top}) Kernel density estimates of the marginal prior $\ln p(A_0|\Hnl)$, (\textit{middle}) marginal posterior $\ln p(A_0|\mathrm{data}, \Hnl)$, and (\textit{bottom}) the associated \lnB~using \Syr's publicly available uniform-mass, small $f_0$ posterior samples.
        We evaluate these at several values of $\log_{10} A_0$.
        These distributions reflect the statistical uncertainty in the exact value of the priors, posteriors, and \lnB~from the finite number of samples.
        By taking the difference between $\ln p(A_0|\Hnl)$ and $\ln p(A_0|\mathrm{data}, \Hnl)$, we obtain the distributions for \lnB.
    }
    \label{figure:syr sddr syr priors}
\end{figure}

In order to compute Bayes Factors, \Syr~uses thermodynamic integration \cite{Nicolas2006, Foremanmackey2013, Vousden2016, Wang2005, Earl2005, Biwar2018} while \LVC~uses the Savage-Dickey Density Ratio (SDDR, see \cite{Dickey1970, Verdinelli1995, Wagenmakers2010} and Appendix \ref{appendix:sddr}).\footnote{\LVC~notes that they cross-checked their SDDR estimates of \lnB~against both nested sampling \cite{skilling2006} and thermodynamic integration.}
In principle, both should yield consistent results.
However, when we apply the SDDR to \Syr's (uniform-mass, small $f_0$) posterior samples, we find $\lnB\approx\approxSyrSampSyrPrior$ whereas \Syr~claims $\lnB = -21$ ($\logtenB = -9.2$) for the same posterior samples.\footnote{\Syr~considers several different mass and $f_0$ priors. For the narrow (broad) $f_0$ prior, they find $\logtenB = -9.2, -6.0$ ($\logtenB = -6.3, -4.7$) for the uniform and Gaussian mass priors, respectively.  They summarize their results by stating that they find $\logtenB < -4$.}

The SDDR provides a convenient way to estimate Bayes Factors between nested models when the posterior is available for the larger model.
In the limit of small $A_0$  (see \LVC's Equation (3) and Appendix \ref{appendix:sddr}), 
\begin{equation}\label{equation:sddr}
    \lim\limits_{A_0\rightarrow 0} \frac{p(A_0|\mathrm{data}, \Hnl)}{p(A_0|\Hnl)} \simeq \frac{p(\mathrm{data}|\Hgr)}{p(\mathrm{data}|\Hnl)} = \frac{1}{B^{pg}_{!pg}}.
\end{equation}
This says that \Bayes~equals the ratio of the marginal distribution of $A_0$ \emph{a priori} to the marginal distribution of $A_0$ \emph{a posteriori} provided that both are evaluated at small $A_0$.
Intuitively, if at some small $A_0$ the probability density is small \emph{a priori} but large \emph{a posteriori}, then the data must favor small $A_0$ and \Bayes~should be small (and vice versa). 

\Syr~finds $\Bayes<10^{-4}$, which implies that, at small $A_0$, their prior density is $ > 10^4$ times smaller than their posterior density.
However, Figures~1--3 in \Syr~show that their prior and posterior densities are similar at small $A_0$, which instead suggests $\Bayes \approx 1$.
More precisely, we find that the SDDR yields $\Bayes \simeq +2.0$ for \Syr's uniform-mass, small $f_0$ prior and similar \Bayes~for their other priors.
Since $\Bayes>1$, formally \Syr's results actually slightly favor the existence of detectable \pg~effects.  However, the preference is not large enough to be significant (see the calculation of the False Alarm Probability in \LVC).

We show this result in more detail in Figure~\ref{figure:syr sddr syr priors}.
Using \Syr's samples, we plot the probability densities of the priors, posteriors, and their ratio (i.e., \Bayes) at several values of small $A_0$.\footnote{To render our kernel density estimation computationally tractable, we select a random subsample of $\approx 5000$ of their posterior samples. Including more samples would decrease the variance of each distribution shown in Figure \ref{figure:syr sddr syr priors}, but we already rule out \Syr's \Bayes~with only 5000 samples.}
Because closed-form expressions for the priors and posteriors are not available, we estimate these from \Syr's public samples.
There are only a finite number of samples available, and the distributions in Figure \ref{figure:syr sddr syr priors} show the uncertainty in our estimates of the prior and posterior at a few values of $\log_{10} A_0$.
These figures were generated using the same code \LVC~used to estimate \lnB in their Figure~1.
For brevity, we focus on \Syr's uniform-mass, small $f_0$ range posteriors (their Figure~2) but find similar results for their other data.
Although the posteriors and priors are not constant as $A_0\rightarrow 0$, what is important is that their ratio, and hence \Bayes, are consistently $\mathcal{O}(1)$.

\section{Examination of \Syr's Implementation of Thermodynamic Integration}
\label{section:thermodynamic integration}

To investigate \Syr's systematic error further, we attempt to repeat their calculation using thermodynamic integration.
Thermodynamic integration makes use of the convenient identity for the evidence $Z_\mathcal{H} = p(d|\mathcal{H})$
\begin{align}
    \frac{d}{d\beta} \ln Z_\mathcal{H}(\beta) & = \frac{d}{d\beta} \ln \int d\theta\, p(d|\theta)^\beta p(\theta|\mathcal{H}) \nonumber \\
                                              & = \frac{\int d\theta \left(\ln p(d|\theta)\right) p(d|\theta)^\beta p(\theta|\mathcal{H})}{\int d\theta\, p(d|\theta)^\beta p(\theta|\mathcal{H})} \nonumber \\
                                              & = \left< \ln p(d|\theta) \right>_{p_\beta(\theta|d;\mathrm{H})}
\end{align}
By running several parallel Markov-Chain Monte Carlo instances, each at a different temperature $T=1/\beta$, one can approximate $d\ln Z/d\beta$ as a function of $\beta$ by computing the average of $\ln p(d|\theta)$ with respect to $p_\beta(\theta|d;\mathcal{H}) \propto p(d|\theta)^\beta p(\theta|\mathcal{H})$.
With enough temperatures, an estimate for the evidence is obtained as
\begin{equation}
    \ln Z_\mathcal{H} = \int\limits_0^1 d\beta \frac{d}{d\beta} \ln Z_\mathcal{H} = \int\limits_0^1 d\beta \left< \ln p(d|\theta) \right>_{p_\beta(\theta|d; \mathrm{H})}
\end{equation}
Typically, the set of temperatures is chosen to optimize sampling through parallel tempering (see, e.g., \cite{Vousden2016, Earl2005}) and the resulting integral is estimated via a trapazoidal approximation.
This procedure is repeated for each model separately, and then differences in the evidences yield Bayes Factors.

\Syr~attempt to implement this approach.
However, as shown by their public data, they only use 3 temperatures, and the smallest inverse temperature used is $\beta\simeq0.25$.
When checking our implementation of the Savage-Dickey Density Ratio against thermodynamic integration on our own data, we only found convergence with at least 12 temperatures.
This means that not only do they poorly resolve the numeric integral, they severely truncate the estimate.
Therefore, the \Syr~result contains large systematic errors.

\Syr~rely upon evidence estimates from previous work \cite{De2018} when estimating Bayes Factors.
Examination of the public data from \cite{De2018} shows that they also used only 3 temperatures and incorrectly truncated the integral.
The Bayes factors originally quoted in \cite{De2018} are therefore also in error (as the erratum in \cite{De2018} acknowledges).

\section{Conclusions}
\label{section:conclusions}

We analyzed the publicly available samples from Reyes \& Brown \cite{Reyes2018}.
Using our own method for calculating Bayes Factors \cite{LVCnltides}, we find that their samples yield $\Bayes\simeq\approxSyrBayes$ and not $\Bayes < 10^{-4}$.
The source of their errors stems from a flawed implementation of thermodynamic integration, which also affected some of the authors' other work (see erratum in \cite{De2018}).
We therefore conclude that, contrary to Reyes \& Brown's claim, their posterior data do not disfavor \pg~mode coupling.

\acknowledgments
The authors thank Steven Reyes and Duncan Brown for making their samples publicly available, and Katerina Chatziioannou, Anuradha Samajdar, Aaron Zimmerman, and the other LVC reviewers for their useful feedback while preparing this note.
R. Essick is supported at the University of Chicago by the Kavli Institute for Cosmological Physics through an endowment from the Kavli Foundation and its founder Fred Kavli.
N. Weinberg was supported in part by NASA grant NNX14AB40G.


\bibliography{refs}

\appendix

\begin{widetext}
\section{Derivation of the Savage-Dickey Density Ratio}
\label{appendix:sddr}

According to Bayes theorem
\begin{equation}
    p(A_0, f_0, n_0, \theta|D; \Hnl) = \frac{1}{p(D|\Hnl)}p(D|A_0, f_0, n_0, \theta; \Hnl)p(A_0, f_0, n_0, \theta|\Hnl),
\end{equation}
where $\theta$ refers to all parameters besides the \pg~parameters and $D$ to the data from GW170817.
We drop the first-order terms in the \pg~parameter Taylor expansions for clarity, but they could be included in a straightforward way.
The marginal posterior distribution for $A_0$ is therefore
\begin{align}
    p(A_0|D; \Hnl) & = \frac{1}{p(D|\Hnl)} \int d\theta df_0 dn_0 p(D|A_0, f_0, n_0, \theta; \Hnl)p(A_0, f_0, n_0, \theta|\Hnl) \nonumber \\
                               & = \frac{1}{p(D|\Hnl)} \int d\theta df_0 dn_0 p(D|A_0, f_0, n_0, \theta; \Hnl)p(f_0, n_0|A_0, \theta; \Hnl) p(\theta|A_0;\Hnl)p(A_0|\Hnl).
\end{align}
Although \Hgr~is not formally contained in \Hnl~for uniform-in-$\log_{10} A_0$ priors, the lower limit of $A_0 = 10^{-10}$ (in both studies) is sufficiently small that \Hgr~is, to a very good approximation, nested in \Hnl.
In particular, at $A_0 = 10^{-10}$, the waveforms of \Hnl~ and \Hgr~match to $> 99.999\%$.
Therefore, in the limit $A_0\rightarrow 10^{-10}$, the likelihood $p(D|A_0, f_0, n_0, \theta; \Hnl)=p(D|\theta; \Hgr)$ and the integral factors.
We then have
\begin{align}
    \lim\limits_{A_0\rightarrow 10^{-10}} \frac{p(A_0|D; \Hnl)}{p(A_0|\Hnl)} & = \frac{1}{p(D|\Hnl)}\lim\limits_{A_0\rightarrow 10^{-10}} \int d\theta df_0 dn_0 p(D|A_0, f_0, n_0, \theta; \Hnl)p(f_0, n_0|A_0, \theta, \Hnl)p(\theta|A_0;\Hnl) \nonumber \\
                                                              & = \frac{1}{p(D|\Hnl)}\lim\limits_{A_0\rightarrow 10^{-10}} \left( \int d\theta p(D|\theta; \Hgr) p(\theta|A_0; \Hnl) \right. \nonumber \\
                                                              & \left. \quad \quad \quad \quad  \quad \quad \quad \quad  \quad \quad \quad \quad  \quad \quad \quad \quad  \quad \quad \quad \quad \times \left[\int df_0 dn_0 p(f_0, n_0|\theta, A_0; \Hnl)\right] \right).
\end{align}
This allows us to integrate away the conditional prior for $f_0$ and $n_0$ and obtain
\begin{align}
    \lim\limits_{A_0\rightarrow 10^{-10}} \frac{p(A_0|D; \Hnl)}{p(A_0|\Hnl)} & = \frac{1}{p(D|\Hnl)} \lim\limits_{A_0\rightarrow 10^{-10}} \int d\theta p(D|\theta; \Hgr) p(\theta|\Hgr) \left[\frac{p(\theta|A_0;\Hnl)}{p(\theta|\Hgr)}\right] \nonumber \\
                                                              & = \frac{p(D|\Hgr)}{p(D|\Hnl)} \lim\limits_{A_0\rightarrow 10^{-10}} \left( \int d\theta p(\theta|D;\Hgr) \frac{p(\theta|A_0;\Hnl)}{p(\theta|\Hgr)} \right).
\end{align}
Thus
\begin{equation}\label{equation:full sddr}
    \lim\limits_{A_0\rightarrow 10^{-10}} \frac{p(A_0|D; \Hnl)}{p(A_0|\Hnl)} = \frac{1}{\Bayes} \lim\limits_{A_0\rightarrow 10^{-10}}\left<\frac{p(\theta|A_0;\Hnl)}{p(\theta|\Hgr)}\right>_{p(\theta|D; \Hgr)},
\end{equation}
where $\left< x \right>_p$ denotes the average of $x$ with respect to the measure defined by $p$.
As we demonstrate in Figure \ref{figure:prior ratio},  $\lim_{A_0\rightarrow 10^{-10}} p(\theta|A_0;\Hnl) \approx p(\theta|\Hgr)$.
The red curves show the conditional prior distributions for component masses ($m_1$, $m_2$), chirp mass ($\mathcal{M}$), and mass ratio ($q=m_2/m_1 \leq 1$) under \Syr's \Hnl~priors for $\log_{10} A_0 \in [-10, -9.9]$.
The blue curves show the distributions for an analogous prior with uniform distributions for $m_1$ and $m_2$ and the same $\mathcal{M}$ cuts but without any requirement on $\delta\phi$, which is \Syr's corresponding \Hgr~prior.
While we see some small differences in the marginal distributions, these are all $\mathcal{O}(1)$.
We therefore expect $\left<p(\theta|A_0;\Hnl)/p(\theta|\Hgr)\right>_{p(\theta|\Hgr)}$ to be a negligible correction and omit it from the main body of this note (Equation (\ref{equation:sddr})).
Therefore, the ratio of the marginal posterior to the marginal prior, when evaluated at sufficiently small $A_0$, yields an accurate an estimate of \Bayes.

\begin{figure}
    \includegraphics[width=\textwidth]{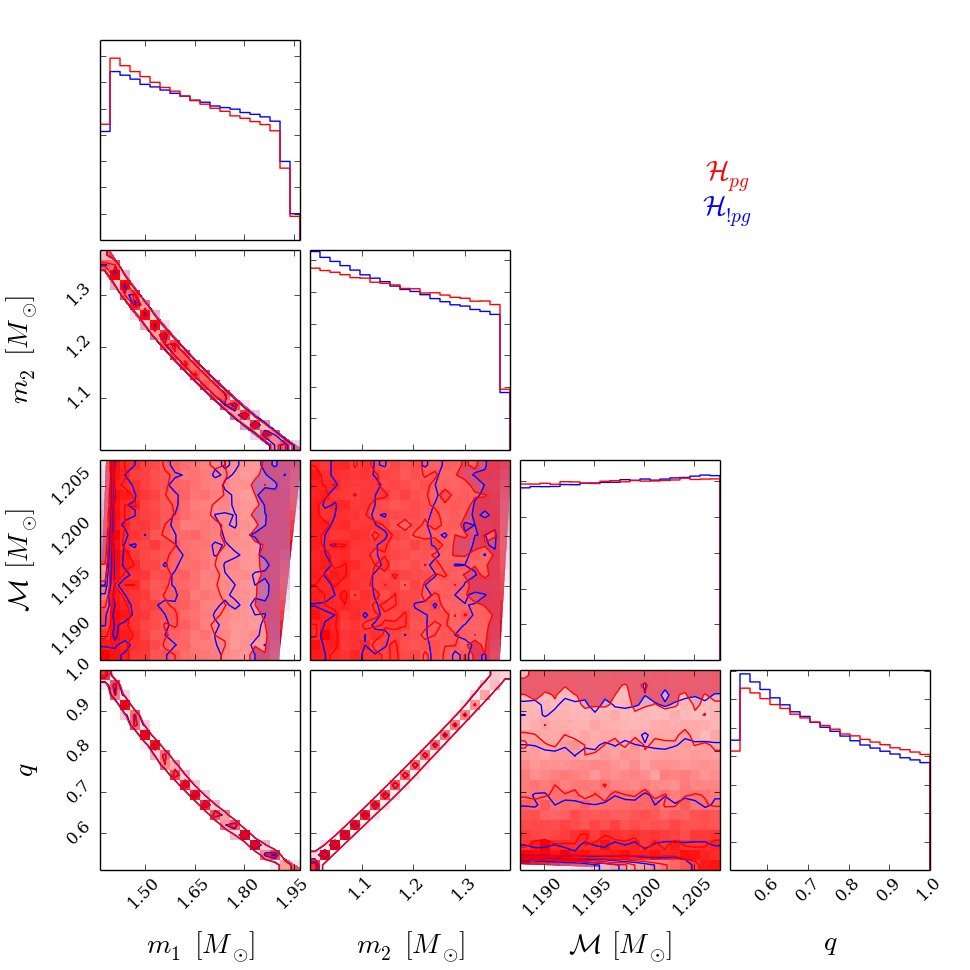}
    \caption{
        Mass priors in \Syr's analysis, which enter as a correction factor in Equation (\ref{equation:full sddr}).
        (\textit{red}) $p(\theta|A_0; \Hnl)$ with $\log_{10} A_0 \in [-10, -9.9]$.
        (\textit{blue}) $p(\theta|\Hgr)$ assuming uniform priors on component masses and the same $\mathcal{M}$ cuts \Syr~uses for \Hnl.
    }
    \label{figure:prior ratio}
\end{figure}

\end{widetext}

\end{document}